# Interference of Hydrogen Atom 2P$_{1/2}$ State in a Field of a Few Small Perturbations

Yu.A. Kucheryaev[1] and Yu.L. Sokolov[2]

Russian Research Center "Kurchatov Institute", 123182 Moscow, Russia

**Abstract.** The interference of hydrogen atom 2P$_{1/2}$ state in a field of a few small overlapping perturbations is considered in view of further applications to experimental data interpretation. On a basis of this model two new experiments are proposed which can clarify some features of Sokolov effect.

**PACS.** 31.30.Jv  Relativistic and quantum electrodynamic effects in atoms and molecules – 39.10.+j  Atomic and molecular beam sources and techniques – 39.20.+q  Atom interferometry

## 1 Introduction

A superposition of 2S$_{1/2}$ (F = 0, F$_z$ = 0) and 2P$_{1/2}$ (F = 1, F$_z$ = 0) states is used as a "tool" for Sokolov effect investigations [1, 2]. During several years the superposition was created by the device with rather strong electric field in which the energy of a perturbation was comparable with energy of the transition ("electrostatic exciter"). The interferogram measured in these experiments (i.e. count rate of $L_\alpha$ quanta emitted by short-lived 2P atoms vs distance between an exciter and analyzer) represented a fading exponent slightly modulated by fading oscillations after subtraction of a background. To estimate the effect magnitude the interferogram was approximated by a function

$$F(L) = a_1 \exp\left(-\frac{L}{R}\right) + a_2 \exp\left(-\frac{L}{2R}\right)\cos\left(\frac{2\pi L}{a_3} + a_4\right) + a_5 , \qquad (1.1)$$

where $L$ is a distance between an exciter and analyzer, $R$ is a range of 2P atom, $a_1...a_5$ are empiric parameters. Such form of $F(L)$ can be justified if it is supposed that between an exciter and investigated perturbation there is a space of size $L$ without any perturbations.

It has become clear during these experiments that it is possible to use a perturbation created by a slit in a grounded metallic plate which follows the collimator as an exciter of a primary superposition instead of an electric field. A scheme of a simplest experiment of such kind is shown in a Fig. 1. Since 1999 till now various modifications of this scheme were used in all experiments on Sokolov effect. In many cases one can approximate the experimental data with admissible accuracy by a function (1.1).

The nature of perturbations created by exciting and analyzing slits is obscure till now. According to the most advanced hypothesis of Kadomtsev [3] they can have rather bizarre form and considerable extension in space. The last means that these perturbations can partially or completely be overlapped depending on $L$ value. Therefore it is not obvious beforehand that the function (1.1) obtained for the case of nonoverlapping perturbations will be suitable in this case as well. The purpose of the following analysis is to examine an evolution of 2P state amplitude in a field of several small perturbations which can excite dipole 2S$_{1/2}$ - 2P$_{1/2}$ transition. It is supposed that each perturbation is localized basically in some limited space, but can be overlapped with other perturbations (particular form of a spatial distribution is not significant). The energy of perturbations is supposed small as compared with the energy of transition, so the 1-st order perturbation theory is applicable with good accuracy.

---

[1] e-mail: 6749.g23@g23.relcom.ru
[2] e-mail: lukich@nfi.kiae.ru

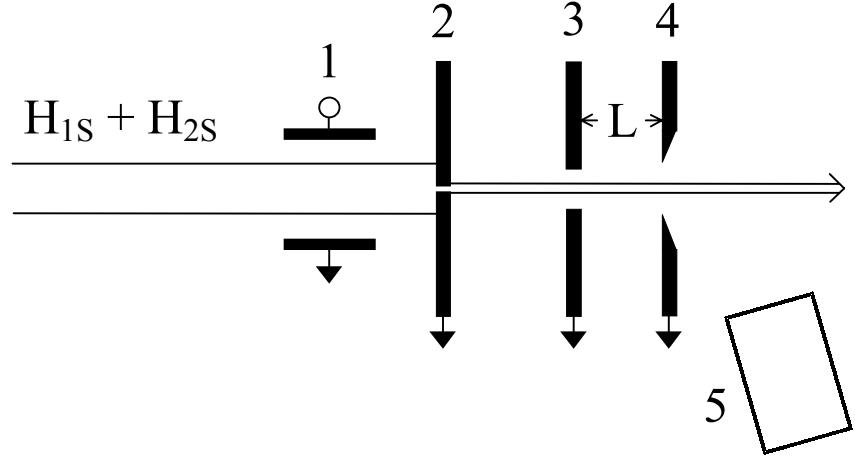

**Fig. 1.** Scheme of the experiment without electrostatic exciter: 1 – quencher of 2S component of atomic beam; 2 – collimator; 3 – exciting slit; 4 – analyzing slit; 5 – $L_\alpha$ detector.

## 2 Approach to the problem

The wave function of 2-level system under consideration at presence of a perturbation can be written as a superposition of nonperturbed functions:

$$\psi(\vec{r}',t) = \sum_m A_m(t)\psi_m(\vec{r}'), \qquad m = s, \, p. \qquad (2.1)$$

Here $\vec{r}'$ is a vector with the origin at the proton, the functions $A_m(t)$ include a phase factor $\exp(-iE_m t/\hbar)$ and in the following are called as "state amplitudes". A substitution of (2.1) in a wave equation

$$i\hbar \frac{\partial \psi}{\partial t} = (\hat{H} + \hat{U})\psi$$

with a nonperturbed Hamiltonian $\hat{H}$ and perturbation operator $\hat{U}$ results in a system of equations for $A_m(t)$

$$\frac{dA_m}{dt} = -i\sum_n (\omega_n \delta_{mn} + u_{mn})A_n, \qquad n = s, \, p, \qquad (2.2)$$

where
$$\omega_n = E_n/\hbar, \qquad u_{mn} = <m|\hat{U}|n>/\hbar.$$

Let us suppose that the perturbation is created by an electric field $\vec{f}$ or some "effective" field that acts the same way as an electric field. This field is stationary in a laboratory reference frame. The measurements of intensity of 2S and 2P atomic beam components are carried out not in the time but in the laboratory space. Let $z$ axis of laboratory coordinate system is directed along atom velocity $\vec{\upsilon}_a$, then atom coordinate is $z = \upsilon_a t$ and one can rewrite (2.2) as

$$\frac{dA_m}{dz} = -i\sum_n (k_n \delta_{mn} + w_{mn}) A_n, \qquad (2.3)$$

where
$$k_n = \omega_n/\upsilon_a, \qquad w_{mn} = <m|\hat{U}|n>/(\hbar \upsilon_a).$$

The nonrelativistic relation between $z$ and $t$ is admissible because $\upsilon_a/c \lesssim 10^{-2}$ in the experimental conditions.

The perturbation operator is $\hat{U} = -\vec{f}\hat{\vec{D}}$, operator of dipole moment is $\hat{\vec{D}} = -e\vec{r}'$. Let us substitute in $\hat{U}$ a field $\vec{f} = f(z)\vec{e}_f$ where $f(z)$ is a field profile, $\vec{e}_f$ is a unit vector in a field direction:

$$\hat{U} = f(z) e \vec{e}_f \vec{r}'.$$

As it is known, the electric field mixs states with opposite parity. Therefore it is possible to write $w_{mn}$ in the form

$$w_{mn} = -q \begin{pmatrix} 0 & 1 \\ 1 & 0 \end{pmatrix}, \qquad (2.4)$$

where $q = Df/(\hbar \upsilon_a)$, $D$ is modulus of a projection of transition dipole moment on the direction $\vec{e}_f$. It is convenient to accept a metastable level $2S_{1/2}$ (F = 0) to be an origin of energy readout, then

$$k_s = 0, \qquad k_p \equiv k = -k_0 - \frac{i}{2R}, \qquad (2.5)$$

where $k_0 = \omega_0/\upsilon_a$, $\omega_0 = 2\pi \nu_0$, $\nu_0$ is $2S_{1/2}$ (F = 0, $F_z$ = 0) - $2P_{1/2}$ (F = 1, $F_z$ = 0) transition frequency equal to 909.89 MHz, $R = \tau \upsilon_a$, $\tau$ = 1.5962 ns is the lifetime of 2P atom determined by its transition to the ground state 1S. The substitution (2.4), (2.5) in (2.3) gives

$$\left. \begin{array}{l} \dfrac{dA_s}{dz} = iqA_p, \\ \dfrac{dA_p}{dz} = iqA_s - ikA_p. \end{array} \right\} \qquad (2.6)$$

Eqns (2.6) can be expressed in dimensionless form if one enters

$$x = k_0 z, \qquad \beta = \frac{1}{2\omega_0 \tau}, \qquad \varphi(x) = \frac{q}{k_0} = \frac{Df}{\hbar \omega_0}. \qquad (2.7)$$

Then

$$\left. \begin{array}{l} \dfrac{dA_s}{dx} = i\varphi A_p, \\ \dfrac{dA_p}{dx} = i\varphi A_s + (i - \beta) A_p. \end{array} \right\} \qquad (2.8)$$

If $|\varphi| \ll 1$ and boundary conditions are $A_s(x_0) = S_0$, $A_p(x_0) = P_0$ ($|P_0| \ll 1$), then only equation for $A_p$ remains from (2.8):

$$\frac{dA_p}{dx} + (\beta - i)A_p = iS_0\varphi .$$

Its solution in a point $x_*$ is

$$A_p(x_*) = \exp[(i-\beta)(x_* - x_0)]\left\{ iS_0 \int_{x_0}^{x_*} \exp[(\beta - i)(x - x_0)]\varphi(x)dx + P_0 \right\}. \qquad (2.9)$$

## 3  Interference in a field of two small perturbations

**3.1** The perturbations which act on atom in the configuration of Fig. 1 are shown schematically in Fig. 2.

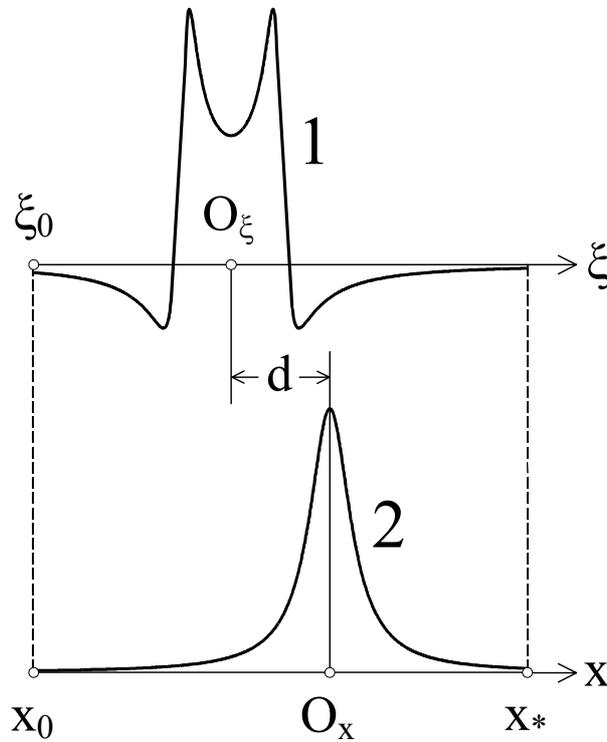

**Fig. 2.** Scheme of the perturbations created by slits of different shape (3 and 4 in Fig. 1): $1 - \varphi_1(\xi)$; $2 - \varphi_2(x)$. Kadomtsev fields are shown in this figure as an example.

For better visualization the $\xi$ axis of the 1-st perturbation coordinate system is biased relatively of $x$ axis of the 2-nd perturbation coordinate system though really they lie on one line. An origin of $\xi$ axis (point $O_\xi$) and origin of $x$ axis (point $O_x$) are chosen somewhere in the area of localization of the corresponding perturbations. The coordinates $\xi_0$ and $x_0$ correspond to output edge of a collimator slit. The $L_\alpha$ quanta are detected from small neighborhood of point $x_*$. Distance $O_\xi O_x$ is equal to $d = d_0 + \ell$ where $d_0 \geq 0$ is constant displacement and $\ell \geq 0$ is changed during experiment from 0 up to $\ell_{max}$. Then the total perturbation in a point $x$ is

$$\varphi(x) = \varphi_1(x + d) + \varphi_2(x) . \qquad (3.1)$$

The substitution (3.1) in (2.9) gives

$$\left.\begin{array}{l} A_p(x_*,d) = iS_0 \exp[(i-\beta)x_*]\{(z_1+\rho_0)\exp[(i-\beta)d] + z_2\}, \\ z_1 = \int_{\xi_0}^{x_*+d} \exp[(\beta-i)x]\varphi_1(x)\,dx, \\ z_2 = \int_{\xi_0-d}^{x_*} \exp[(\beta-i)x]\varphi_2(x)\,dx, \\ \rho_0 = -\dfrac{iP_0}{S_0}\exp[(\beta-i)\xi_0]. \end{array}\right\} \quad (3.2)$$

Square of $A_p(x_*,d)$ modulus as a function of $d$ represents the most general expression for an interferogram of a 2P state within the frame of model under consideration. It can be written as

$$|A_p|^2(d) = \\ = |S_0|^2 \exp(-2\beta x_*)[|z_1+\rho_0|^2 e^{-2\beta d} + 2|z_1+\rho_0||z_2|e^{-\beta d}\cos(d+\alpha_1-\alpha_2) + |z_2|^2], \quad (3.3)$$

where $\alpha_1$, $\alpha_2$ are phases of $z_1$, $z_2$. The Eqn (3.3) is similar to "standard" function (1.1) only outwardly because $z_1$ and $z_2$ depend on $d$ in general case. This dependence can appear weak if width of distributions $\varphi_1(\xi)$ and $\varphi_2(x)$ is small as compared with size of an interaction region $x_* - x_0$. Let us introduce "principal values" $z_{10}$, $z_{20}$ of perturbation integrals $z_1$, $z_2$:

$$z_1 + \rho_0 = z_{10} + \Delta_1, \qquad z_2 = z_{20} + \Delta_2, \quad (3.4)$$

where

$$\left.\begin{array}{l} z_{10} = \int_{\xi_0}^{x_*} \exp[(\beta-i)x]\varphi_1(x)\,dx, \\ z_{20} = \int_{\xi_0}^{x_*} \exp[(\beta-i)x]\varphi_2(x)\,dx. \end{array}\right\} \quad (3.5)$$

Now only the corrections depend on $d$:

$$\left.\begin{array}{l} \Delta_1 = \int_{x_*}^{x_*+d} \exp[(\beta-i)x]\varphi_1(x)\,dx + \rho_0, \\ \Delta_2 = \int_{\xi_0-d}^{\xi_0} \exp[(\beta-i)x]\varphi_2(x)\,dx. \end{array}\right\} \quad (3.6)$$

Integrals in (3.6) are calculated in regions which are far from functions $\varphi_1(\xi)$ and $\varphi_2(x)$ localization areas. The value of $\rho_0$ contains small factor $\exp(-\beta|\xi_0|)$. So, supposing $|\Delta_n|^2 \ll |z_{n0}|^2$, one has from (3.4)

$$\left.\begin{array}{l} |z_1+\rho_0|^2 \approx |z_{10}|^2 + z_{10}\Delta_1^* + z_{10}^*\Delta_1, \\ |z_2|^2 \approx |z_{20}|^2 + z_{20}\Delta_2^* + z_{20}^*\Delta_2. \end{array}\right\} \quad (3.7)$$

Let us replace the integrals in (3.6) by their approximate expressions limited to linear on $d$ terms. Then

$$\Delta_1 \approx \Delta_{10} + \rho_0, \qquad \Delta_2 \approx \Delta_{20}, \qquad (3.8)$$

$$\left.\begin{array}{l}\Delta_{10} = \exp[(\beta - i)x_*]\varphi_1(x_*)d, \\ \Delta_{20} = \exp[(\beta - i)\xi_0]\varphi_2(\xi_0)d.\end{array}\right\} \qquad (3.9)$$

After straightforward calculations one finds from (3.7) ... (3.9)

$$\left.\begin{array}{l}|z_1 + \rho_0|^2 \approx |z_{10}|^2(1 + 2\varepsilon_0 + 2\varepsilon_1 d), \\ |z_2|^2 \approx |z_{20}|^2(1 + 2\varepsilon_2 d),\end{array}\right\} \qquad (3.10)$$

where

$$\left.\begin{array}{l}\varepsilon_0 = |\rho_0|\cos(\alpha_{10} - \arg\rho_0)/|z_{10}|, \\ \varepsilon_1 = \exp(\beta x_*)\varphi_1(x_*)\cos(\alpha_{10} + x_*)/|z_{10}|, \\ \varepsilon_2 = \exp(\beta \xi_0)\varphi_2(\xi_0)\cos(\alpha_{20} + \xi_0)/|z_{20}|,\end{array}\right\} \qquad (3.11)$$

$\alpha_{10}, \alpha_{20}$ are phases of $z_{10}, z_{20}$.

The initial amplitude $P_0$ is produced by the interaction of 2S atoms beam with a collimator slit. It is difficult to estimate the value of $P_0$ but it is visible from (3.10) that its influence can be neglected if the condition $2\varepsilon_0 \ll 1$ is fulfilled. Taking into account (3.11) and (3.2) at $|P_0|/|S_0| \leq 1$ one can estimate a necessary distance between the collimator 2 and slit 3 (see Fig. 1):

$$|\xi_0| \gg \frac{1}{\beta}\ln\frac{2}{|z_{10}|}. \qquad (3.12)$$

Supposing that (3.12) takes place and neglecting terms including $\varepsilon_i \varepsilon_k$ one can rewrite (3.3) as

$$|A_p|^2 \exp(2\beta x_*)/|S_0|^2 \approx$$
$$\approx |z_{10}|^2(1+2\varepsilon_1 d)e^{-2\beta d} + 2|z_{10}||z_{20}|[1+(\varepsilon_1+\varepsilon_2)d]e^{-\beta d}\cos(d+\alpha_{10}-\alpha_{20}) + |z_{20}|^2(1+2\varepsilon_2 d).$$

The count rate of $L_\alpha$ quanta in the signal detector is proportional to $|A_p|^2$. Accounting for the relation $d = d_0 + \ell$ the count rate can be presented as

$$\left.\begin{array}{l}I_1(\ell) \approx K\{|z_{10}|^2 \exp(-2\beta d_0)[1 + 2\varepsilon_1(d_0 + \ell)]\exp(-2\beta\ell) + \\ + 2|z_{10}||z_{20}|\exp(-\beta d_0)[1 + (\varepsilon_1 + \varepsilon_2)(d_0 + \ell)]\exp(-\beta\ell)\cos(\ell + d_0 + \alpha_{10} - \alpha_{20}) + \\ + |z_{20}|^2[1 + 2\varepsilon_2(d_0 + \ell)]\},\end{array}\right\} \qquad (3.13)$$

where the factor $K$ includes $|S_0|^2 \exp(-2\beta x_*)$, 2S atoms beam intensity, and an efficiency of quanta collection and count as well. Eqn (3.13) can be considered as approximate expression for a calculated interferogram in case of two small overlapping perturbations.

Let us approximate a measured interferogram $Y(\ell)$ by a function of (3.13) type:

$$F_1(\ell) = a_1[1 + 2(a_5 + a_6\ell)]\exp(-2\beta\ell) +$$
$$+ a_2[1 + a_5 + a_7 + (a_6 + a_8)\ell]\exp(-\beta\ell)\cos(\ell + a_3) + a_4[1 + 2(a_7 + a_8\ell)]. \quad (3.14)$$

The relations between empiric parameters and model parameters follow from comparison of (3.14) and (3.13):

$$\begin{aligned}
a_1 &= K|z_{10}|^2 \exp(-2\beta d_0), \\
a_2 &= 2K|z_{10}||z_{20}|\exp(-\beta d_0), \\
a_3 &= d_0 + \alpha_{10} - \alpha_{20}, \\
a_4 &= K|z_{20}|^2, \\
a_5 &= \varepsilon_1 d_0, \\
a_6 &= \varepsilon_1, \\
a_7 &= \varepsilon_2 d_0, \\
a_8 &= \varepsilon_2.
\end{aligned} \quad (3.15)$$

Eight equations for 7 model parameters and parameter $K$ are available. But these equations are such that it is possible to find only 3 model parameters and ratio $|z_{10}|/|z_{20}|$:

$$\varepsilon_1 = a_6, \quad \varepsilon_2 = a_8, \quad d_0 = \frac{a_5 - a_7}{a_6 - a_8},$$
$$\frac{|z_{10}|}{|z_{20}|} = \frac{2a_1}{a_2}\exp(\beta d_0).$$

The relations following from (3.15) can serve for an estimation of approximation quality:

$$a_4 = \frac{a_2^2}{4a_1}, \quad \frac{a_5}{a_6} = \frac{a_7}{a_8}.$$

However this approximation can appear impossible in practice because of inevitable fluctuations of the signal and because of existence of background component which arises together with 2S atoms beam and cannot be measured in experiment.

**3.2** If the distributions $\varphi_1(\xi)$ and $\varphi_2(x)$ appear narrow enough (i.e. conditions $\varepsilon_1 d_{max}$, $\varepsilon_2 d_{max} \ll 1$ are fulfilled) the calculated count rate vs $\ell$ has a form of a "standard" interferogram:

$$I_0(\ell) \approx a_{10}\exp(-2\beta\ell) + a_{20}\exp(-\beta\ell)\cos(\ell + a_{30}) + a_{40}, \quad (3.16)$$

where

$$\begin{aligned}
a_{10} &= K|z_{10}|^2 \exp(-2\beta d_0), \\
a_{20} &= 2K|z_{10}||z_{20}|\exp(-\beta d_0), \\
a_{30} &= d_0 + \alpha_{10} - \alpha_{20}, \\
a_{40} &= K|z_{20}|^2.
\end{aligned} \quad (3.17)$$

Only three of four parameters $a_{10}...a_{40}$ are independent. "Pedestal" $a_{40}$ can be expressed through $a_{10}$ and $a_{20}$:

$$p_0 \equiv a_{40} = \frac{a_{20}^2}{4a_{10}}.$$

Let us approximate a measured interferogram by a function of (3.16) type:

$$F_0(\ell) = a_1 \exp(-2\beta\ell) + a_2 \exp(-\beta\ell)\cos(\ell + a_3) + p + b, \qquad (3.18)$$

where $p = a_2^2/(4a_1)$ and parameter $b$ accounts before-mentioned background component. One finds from (3.16)…(3.18):

$$\begin{aligned} a_1 &= K|z_{10}|^2 \exp(-2\beta d_0), \\ a_2 &= 2K|z_{10}||z_{20}|\exp(-\beta d_0), \\ a_3 &= d_0 + \alpha_1 - \alpha_2. \end{aligned} \qquad (3.19)$$

**3.3** Thus, if a model is available in which the distributions $\varphi_1(\xi)$, $\varphi_2(x)$ are narrow enough and the integral characteristics $|z_{10}|$, $|z_{20}|$ are computed, it is possible to find from (3.17) a calculated value of "percentage modulation of an interferogram"

$$\mu_{calc} \equiv \frac{a_{20}}{a_{10}} = 2\frac{|z_{20}|}{|z_{10}|}\exp(\beta d_0). \qquad (3.20)$$

Comparison of $\mu_{calc}$ with $\mu_{exp} = a_2/a_1$ obtained from experiment allows to appreciate suitability of this model for the description of observed effect. It is clear from (3.19) that having obtained a set of parameters of the function which fits a measured interferogram, but not having any model of "slit" interaction, one can conclude nothing about value and shape of investigated perturbations.

**3.4** Let us suppose that the perturbing influence of a slit on the atom flying through it has "regular" nature, i.e. is determined by such factors which can be controlled during experiment and can be reproduced from run to run (e.g. material of a plate and geometry of a slit, and, probably, temperature of a specimen and atom velocity as well). Hypothetical Kadomtsev fields can serve as an example of "regular" perturbations. If slits 3 and 4 in Fig. 1 are made identical, it is possible to expect that they will create identical perturbations $\varphi_1(\xi)$ and $\varphi_2(x)$. It follows from the above considerations that in this case

$$\mu_{calc} = 2\exp(\beta d_0) \qquad (3.21)$$

irrespective of particular $\varphi_1(\xi)$ and $\varphi_2(x)$ shapes. If $\mu_{exp}$ differs essentially from value (3.21) it means most likely that the real perturbations have another nature (for example, are created by fields of charged dielectric films on electrode surfaces). This case is discussed in more detail in Section 5.

# 4 Additional electric field

**4.1** Let the voltage $U$ is applied between two identical electrodes 3 and 4 in Fig. 1. The perturbations which act on the atom in this case are shown schematically in Fig. 3.

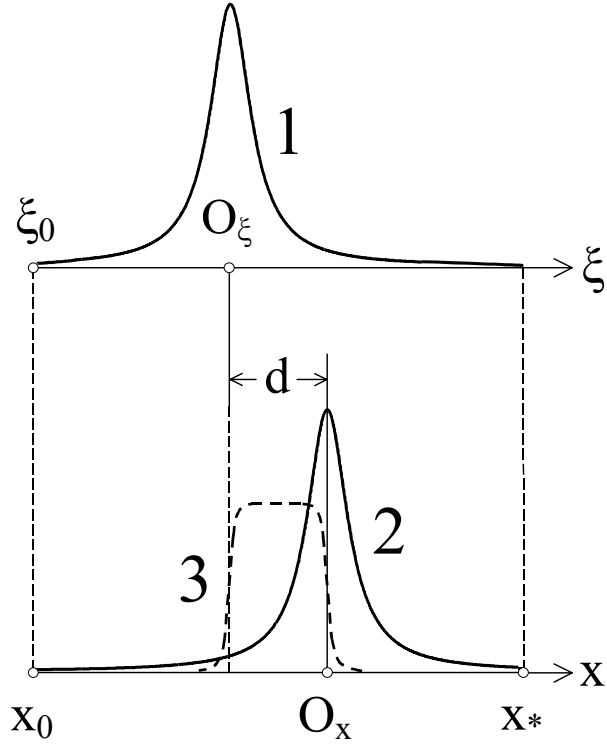

**Fig. 3.** Scheme of the perturbations created by identical slits with additional electric field applied between them: $1 - \varphi_1(\xi)$; $2 - \varphi_2(x)$; $3 - \varphi_e(x, d)$.

Here $\varphi_e(x, d)$ represents a perturbation by an electric field which is proportional to the voltage $U$. Let us suppose also that the distributions $\varphi_1(\xi)$, $\varphi_2(x)$ are narrow enough and that the condition (3.12) is fulfilled, therefore influence of an initial amplitude $P_0$ can be neglected. Allowing in (2.9) $P_0 = 0$ and

$$\varphi(x) = \varphi_1(x + d) + \varphi_2(x) + \varphi_e(x, d),$$

one obtains by the analogy with (3.2)

$$\left. \begin{array}{l} A_p(x_*, d) = iS_0 \exp[(i - \beta)x_*]\{z_1 \exp[(i - \beta)d] + z_2 + z_e\}, \\[6pt] z_1 = \displaystyle\int_{\xi_0}^{x_* + d} \exp[(\beta - i)x]\varphi_1(x)\,dx, \\[6pt] z_2 = \displaystyle\int_{\xi_0 - d}^{x_*} \exp[(\beta - i)x]\varphi_2(x)\,dx, \\[6pt] z_e = \displaystyle\int_{\xi_0 - d}^{x_*} \exp[(\beta - i)x]\varphi_e(x, d)\,dx. \end{array} \right\}$$

**4.2** The additional electric field can be written as

$$f_e(x, d) = \frac{U}{\lambda(x, d)}. \tag{4.1}$$

Let us substitute (4.1) in (2.7):

$$\varphi_e(x, d) = \frac{D f_e(x, d)}{\hbar \omega_0} = \frac{DU}{\hbar \omega_0 \lambda(x, d)} = \frac{U}{U_0} \psi(x, d),$$

$$U_0 \equiv \frac{\hbar \upsilon_a}{D}, \quad \psi(x, d) \equiv \frac{1}{k_0 \lambda(x, d)}.$$

Then $z_e = u\zeta$ where

$$u \equiv \frac{U}{U_0}, \quad \zeta = \int_{\xi_0 - d}^{x_*} \exp[(\beta - i)x] \psi(x, d) \, dx.$$

Count rate of $L_\alpha$ quanta in the signal detector is proportional to $|A_p(x_*, d, u)|^2$. It can be presented as

$$I(u) = b_0 + b_1 u + b_2 u^2, \tag{4.2}$$

where $b_0$, $b_1$, $b_2$ depend on $x_*$ and $d$. If distributions $\varphi_1(\xi)$ and $\varphi_2(x)$ are narrow enough the perturbation integrals can be approximated by their "principal values". The values $z_{10}$ and $z_{20}$ are determined by Eqns (3.5),

$$\zeta_0 = \int_{\xi_0}^{x_*} \exp[(\beta - i)x] \psi(x, d) \, dx. \tag{4.3}$$

Then

$$\left.\begin{array}{l}
b_0 = K\left[|z_{10}|^2 \exp(-2\beta d) + 2|z_{10}||z_{20}|\exp(-\beta d)\cos(d + \alpha_{10} - \alpha_{20}) + |z_{20}|^2\right], \\
b_1 = 2K|\zeta_0|[|z_{10}|\exp(-\beta d)\cos(d + \alpha_{10} - \alpha_{e0}) + |z_{20}|\cos(\alpha_{20} - \alpha_{e0})], \\
b_2 = K|\zeta_0|^2,
\end{array}\right\} \tag{4.4}$$

where $\alpha_{e0}$ is phase of $\zeta_0$, the factor $K$ has the same sense as in (3.13).

If one accepts that in the experiment with two identical slits $z_{10} = z_{20}$ then $|z_{10}| = |z_{20}| = |z|$ and $\alpha_{10} = \alpha_{20}$. Let us measure $I(u)$ dependence at fixed $x_*$, $d$ and approximate it by function (4.2). It follows from (4.4) that

$$\frac{b_0}{b_2} = |z|^2 G(d), \tag{4.5}$$

$$G(d) = [1 + 2\exp(-\beta d)\cos d + \exp(-2\beta d)]/|\zeta_0|^2. \tag{4.6}$$

The calculation of function $|\zeta_0(d)|^2$ is reduced to the solution of an electrostatic problem with known electrodes geometry. For example, if the distance between electrodes is great as compared with the width of slits, then

$$\psi(x, d) \approx \begin{cases} 0 & \text{at } x < -d, \\ \dfrac{1}{d} & \text{at } -d \leq x \leq 0, \\ 0 & \text{at } x > 0. \end{cases} \qquad (4.7)$$

It follows from (4.3) and (4.6) that

$$|\zeta_0|^2 = \frac{1 - 2\exp(-\beta d)\cos d + \exp(-2\beta d)}{(1 + \beta^2)d^2}, \qquad (4.8)$$

$$G(d) \approx d^2 \frac{1 + 2\exp(-\beta d)\cos d + \exp(-2\beta d)}{1 - 2\exp(-\beta d)\cos d + \exp(-2\beta d)}.$$

Thus, the measurement of $I(u)$ dependence at fixed $x_*$ and $d$ in experiment with two identical slits allows in principle to find a modulus of an integral of a perturbation created by each slit.

**4.3** However, it is impossible to exclude beforehand that field of charged dielectric films (FCF) exists between seeing each other surfaces of electrodes 3 and 4 (Fig. 1) along with the field created by applied from outside ("external") voltage $U_{ext}$, provided such films can cover surfaces of the electrodes. The FCF value and distribution in space depends on distribution of density of charges on the surface of films. To show that FCF can variously influence results of measurements, two idealized situations are considered, though the real distribution of density of charges can appear more complex.

- If the charges are concentrated near edges of slits, then the fields created by them are localized inside and in the proximate neighborhood of slits. In the absence of any long-range interactions of the atom with metal surface (for example, if the Kadomtsev hypothesis is wrong) just these "local" FCF can create a perturbations $\varphi_1$ and $\varphi_2$ which result in an interference of a 2P state as shown in Section 3.
- Thickness of the dielectric film and density of charges on it remain constant on significant part of electrode surface, starting from a slit edge up to distances greater than maximum distance between electrodes. Such film is equivalent to a plane capacitor charged up to some voltage. A similar capacitor though with other voltage can exist on the other electrode too. If both electrodes are grounded, the sum of capacitor voltages taken with their signs results in some "internal" voltage $U_{int}$ and a field corresponding to it. One can call this FCF "global" in the distinction of the previous case because this FCF is distributed in the space approximately the same way as a field created by "external" voltage $U_{ext}$. In absence of the perturbations $\varphi_1$ and $\varphi_2$ localized near the slits, global FCF can also create $L_\alpha$ quanta count rate oscillations at $d$ variation, though their shape is different from "standard" interferogram. If one substitutes $u = u_i \equiv U_{int}/U_0$ and $|z_{10}| = |z_{20}| = 0$, it is visible from (4.2) and (4.4) that

$$I(d) = K u_i^2 |\zeta_0(d)|^2.$$

For example, in case of (4.7), (4.8)

$$I(d) \approx K u_i^2 \frac{1 - 2\exp(-\beta d)\cos d + \exp(-2\beta d)}{d^2}.$$

At presence of perturbations $\varphi_1$ and $\varphi_2$ with rather narrow spatial distributions global FCF results in deviations of an interferogram from the "standard" function (3.18). Using Eqns (4.2) and (4.4) again one can obtain:

$$I(d) = K\{|z_{10}|^2 \exp(-2\beta d) + 2|z_{10}||z_{20}|\exp(-\beta d)[\cos(d + \alpha_{10} - \alpha_{20}) +$$
$$+ \theta(d)\cos(d + \alpha_{10} - \alpha_{e0})] + |z_{20}|^2[1 + 2\theta(d)\cos(\alpha_{20} - \alpha_{e0}) + \theta^2(d)]\},$$
$$\theta(d) = \frac{u_i|\zeta_0(d)|}{|z_{20}|}.$$

It is visible that the interferogram becomes "standard" if $\theta(d) \ll 1$, i.e. the perturbation created by "internal" voltage $U_{int}$ should be small as compared with perturbation created by the 2-nd slit.

It is possible to find global FCF from experiment. Figuring in Subsection 4.2 value $U$ can be presented as $U = U_{ext} + U_{int}$ and therefore

$$u = u_e + u_i, \quad \text{where} \quad u_e = U_{ext}/U_0, \quad u_i = U_{int}/U_0. \tag{4.9}$$

Let us substitute $u$ from (4.9) in (4.2):

$$I_{exp}(u_e) = c_0 + c_1 u_e + c_2 u_e^2,$$
$$c_0 = b_0 + b_1 u_i + b_2 u_i^2,$$
$$c_1 = b_1 + 2b_2 u_i, \tag{4.10}$$
$$c_2 = b_2.$$

Let us perform two measurements $I_{exp}(u_e)$ at two different values of $d$. Supposing that $u_i$ and $|z|^2$ are constant in these runs and substituting $b_0$, $b_2$ from (4.10) in (4.5), one can obtain a set of equations

$$\frac{c_0^{(1)}}{c_2^{(1)}} - \frac{c_1^{(1)}}{c_2^{(1)}} u_i + u_i^2 = |z|^2 G(d_1),$$
$$\frac{c_0^{(2)}}{c_2^{(2)}} - \frac{c_1^{(2)}}{c_2^{(2)}} u_i + u_i^2 = |z|^2 G(d_2),$$

which allows to find $u_i$ and $|z|^2$ from experimental data.

## 5 Discussion

Thus, two experiments are proposed: in the 1-st one (Section 3) an interferogram of a 2P state arising as a result of 2S atoms flight through two identical grounded slits is measured; in the 2-nd one (Section 4) intensity of $L_\alpha$ radiation vs voltage between corresponding electrodes at fixed distance between them is measured. The comparison of results of these experiments can promote clarification of the nature of investigated effect. Let us consider some of possible situations.

**5.1** The measured interferogram is not reproduced from run to run, though the conditions controlled during experiment remain constant. It is obvious that main efforts in this case should be directed to search of the factors which essentially influence on measurements data. One of such factors can be FCF which was not controlled in these experiments till now. If after the 1-st experiment to implement the 2-nd one at two different distances between electrodes, it is possible to estimate global FCF value. Since the surface density of charges can depend on intensity of atoms flux passing through slits, it is expedient to repeat both experiments several times at different values of intensity but identical other conditions. The considerable and unstable FCF value indicates necessity of substantial improvement of vacuum conditions in the experiment.

**5.2** The interferogram measured under constant conditions of experiment is quite well reproduced from run to run, but its approximation by a "standard" function (3.18) is not satisfactory. As shown in Subsection 4.3, a reason of such distortion of an interferogram can be global FCF if one admits that there is a rather stable process of dielectric films formation and charging. This assumption can be tested through the 2-nd experiment as well as in the previous case.

**5.3** The measured interferogram is well approximated by a "standard" function (3.18), but the obtained percentage modulation of interferogram significantly differs from expected value (3.21). One can assert that global FCF is absent or negligible in this case (certainly it does not mean that dielectric films on electrodes surfaces are absent). Therefore the 2-nd experiment at four different $d$ values with usage of Eqns (4.2) … (4.4) allows to find modules and phases of perturbation integrals $z_{10}$ and $z_{20}$. The additional information can be given by measurement of $z_{10}$ and $z_{20}$ at different values of atomic beam intensity.

The interpretation of these observations depends on the obtained results. According to Kadomtsev hypothesis, specimens manufactured of an identical material with identical geometry of slits create identical fields if atom velocity is constant. Therefore $|z_{10}| = |z_{20}|$ and the percentage modulation of an interferogram is determined by Eqn (3.21). Let us suppose that the dielectric film on a specimen surface influences on a value of Kadomtsev field too, even if there are no charges on it. Then the values $|z_{10}|$ and $|z_{20}|$ can appear different, but the substitution of them in Eqn (3.20) should give the percentage modulation equal to obtained in the 1-st experiment because the external electric field which is used in the 2-nd experiment can not influence on a Kadomtsev field. If such conformity is not presented, the explanation of observed effect by existence of local FCF seems more plausible, especially if the percentage modulation varies from run to run or depends on intensity of atomic beam. In the 1-st experiment, when the external electric field is absent, the edges of two identical slits can be charged up to different values. It explains why obtained in this experiment percentage modulation of an interferogram differs from value (3.21). The electric field which is used in the 2-nd experiment can change conditions of charge accumulation on edges and therefore can change $z_{10}$ and $z_{20}$ relative to those values which they had in the 1-st experiment. So $\mu_{exp}$ values obtained by two different ways do not coincide. However, if both considered mechanisms of perturbation act jointly, the estimation of their relative contributions seems to be a very difficult problem.

**5.4** The repetition of the 1-st experiment under constant conditions steadily reproduces a "standard" interferogram with percentage modulation close to value (3.21). If the 2-nd experiment performed under the same conditions gives equal values $|z_{10}|$ and $|z_{20}|$ irrespective of atomic beam intensity, it is important proof in favour of Kadomtsev hypothesis. The additional confirmation of this hypothesis validity can be obtained if the repetition of both experiments with two identical slits of other geometry leads to similar results, though with other values $|z_{10}| = |z_{20}|$. One can compute a modulus of perturbation integral, provided Kadomtsev field for one slit is known. Difference between the calculated value and experimental values $|z_{10}| = |z_{20}|$ can be interpreted as effect of dielectric film on a specimen surface. But if the experiments mentioned in this Subsection give other results (for example, $|z_{10}|$ and $|z_{20}|$ obtained in the 2-nd experiment appear different), then in looking for an explanation of observed effects it is necessary to have in mind a possibility of local FCF existence as well.

The discussed in this Section experiments are scheduled to implement at the nearest time on "Pamir" device of RRC "Kurchatov Institute".

The authors are grateful to V.G. Pal'chikov and A.D. Panov for fruitful discussions and valuable advices.

The work was supported by the Russian MINPROMNAUKA (grant 01-53) and by the Russian Foundation for Basic Research (grant RFBR 01-02-16516).